\documentclass{article}
\usepackage{graphicx}\graphicspath{./graph/}%
\usepackage{amssymb,amsmath}
\usepackage{graphicx}\graphicspath{./graph/}%
\usepackage{dcolumn}
\usepackage{bm}
\usepackage{color}
\usepackage{verbatim}
\usepackage{amsfonts}
\usepackage{latexsym}
\usepackage{epstopdf}
\newcommand\pt{\partial}

\newcommand\pr{^\prime}
\newcommand\be{\begin{equation}}
\newcommand\ee{\end{equation}}
\newcommand\bea{\begin{eqnarray}}
\newcommand\eea{\end{eqnarray}}

\newcommand{\Imp}{{\rm Im}~}
\newcommand{\Rep}{{\rm Re}~}
\begin{document}

\title{Coherent Synchrotron Radiation in Whispering Gallery Modes: Theory and Evidence}
\author{Robert Warnock \\ SLAC National Accelerator Laboratory\\
Stanford University, Menlo Park, California 94025, USA\\ and\\
Department of Mathematics and Statistics, University of New Mexico\\
Albuquerque, New Mexico 87131, USA\\ \\
e-mail: warnock@slac.stanford.edu}
\date{}
 \maketitle
\footnote{Invited paper at 5th ICFA Microbunching Instability Workshop, Pohang, South Korea,\\ May 8-10, 2013.}
\begin{abstract}
\large
Theory predicts that Coherent Synchrotron Radiation (CSR) in electron storage rings should appear in whispering gallery modes. In an idealized model these are resonances of the vacuum chamber that are characterized by their high frequencies and concentration of the field near the outer wall of the chamber.
 The resonant modes imply a series of sharp peaks in the frequency spectrum of CSR, and very long wake fields which lead to interbunch communication.
 Theory and experimental evidence for this behavior will be reviewed.
 \end{abstract}
\large
\section{Introduction}
\begin{figure}
\begin{minipage}[b]{.46\linewidth}
\centering\includegraphics[width=40mm]{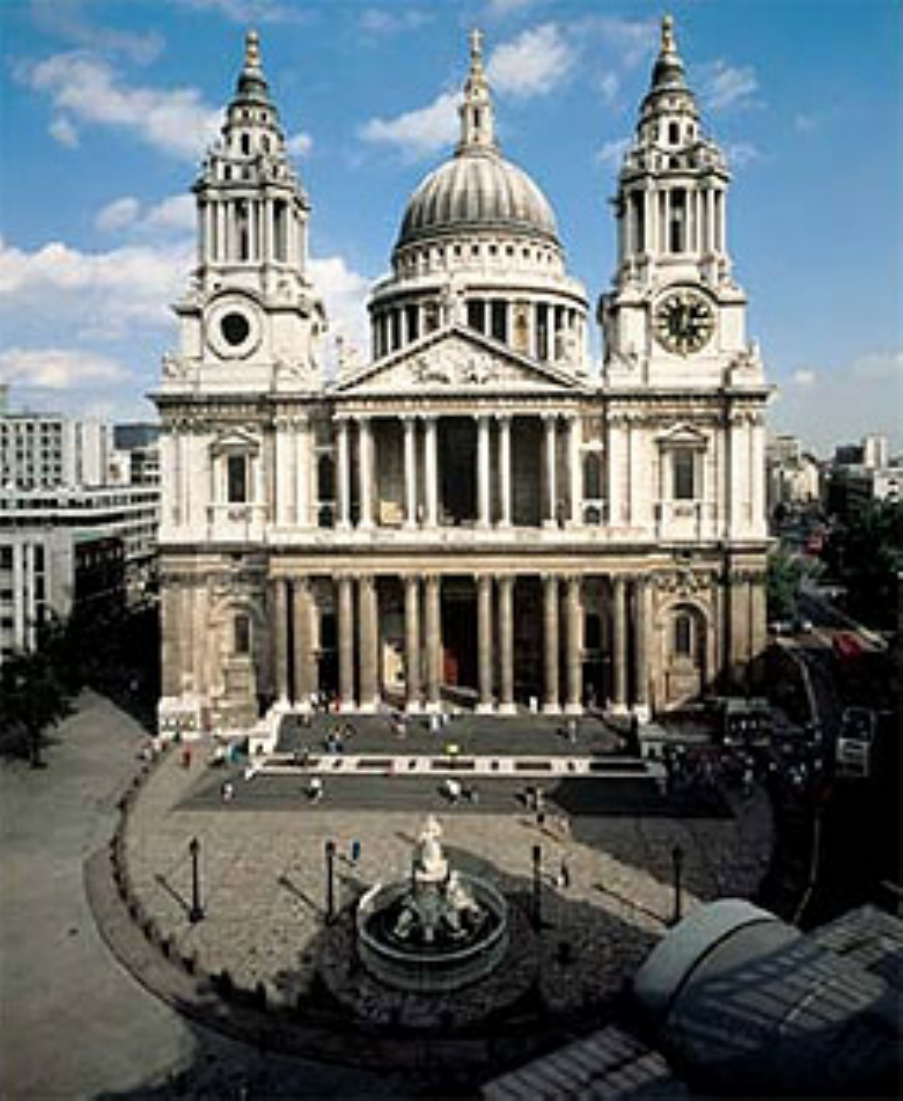}
\caption{St. Paul's Cathedral; Christopher Wren, architect}
\label{stpauls}
\end{minipage}
\begin{minipage}[b]{.46\linewidth}

\centering\includegraphics[width=40mm,]{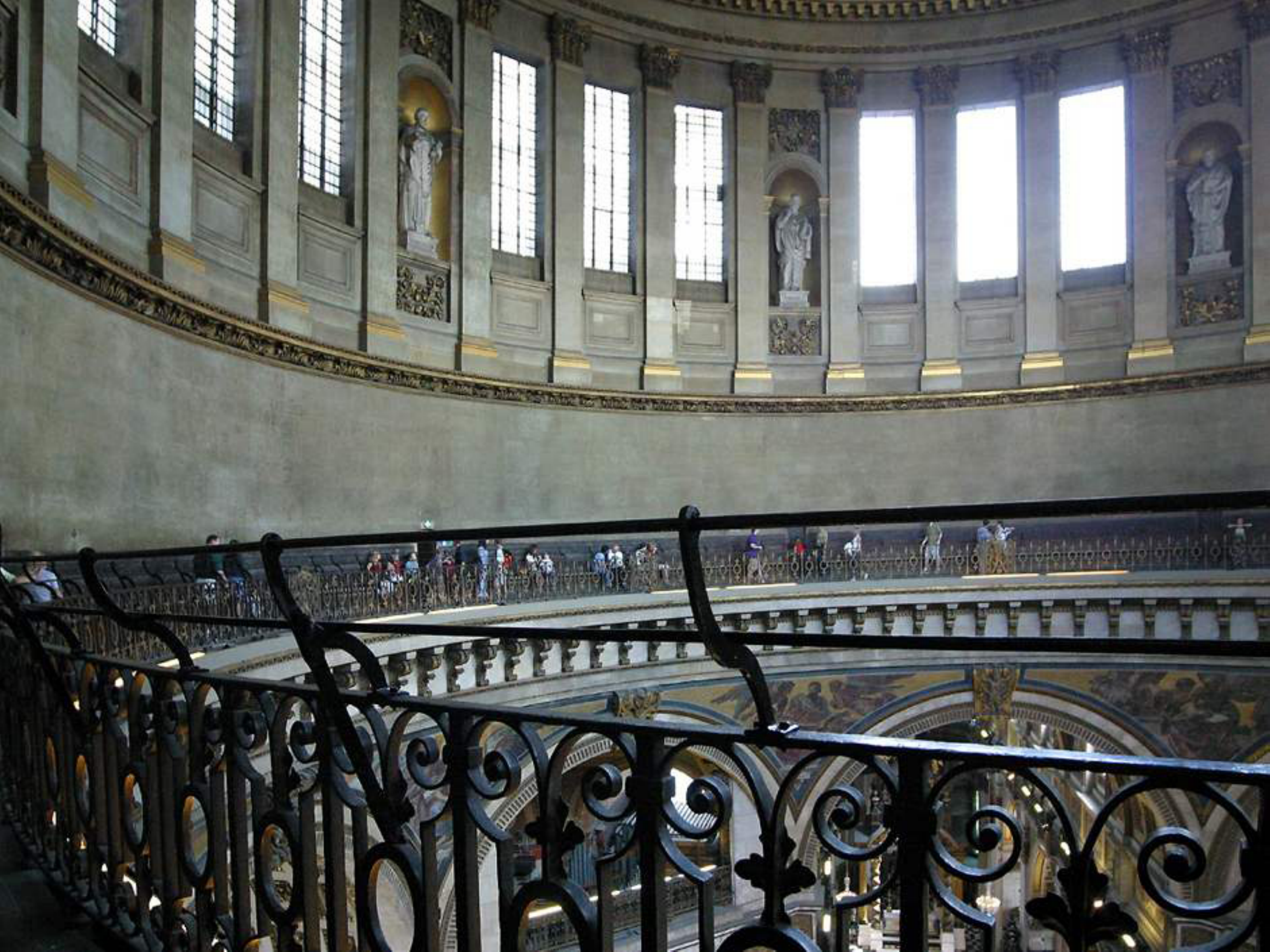}
\vskip 1.2cm
\caption{Whispering Gallery}
\label{gallery}
\end{minipage}
\end{figure}

Whispering galleries in certain public buildings have been a curiosity for centuries. In a building with a cylindrical wall,
one person whispering in a tangential direction near the wall can be heard clearly by another person at a remote location near the wall.
Lord Rayleigh \cite{rayleigh} was inspired to study this phenomenon at St. Paul's Cathedral in London (Fig.\ref{stpauls}), where
there is a famous whispering gallery running near the base of the huge dome (Fig.\ref{gallery}).  George Biddell Airy,  Astronomer Royal at the time of Rayleigh's study, thought the effect had to do with a concentration of an echo in which the symmetry of the dome played a role.  Rayleigh raised doubts about this view since it would require speaker and auditor to be at opposite sides of the gallery, contrary to observations. As it turned out later, Airy functions play a role in a correct description of the effect, since they provide approximations to high order Bessel functions.

One need not be Christopher Wren to design a whispering gallery. I  observed the effect in a distinctly incomparable setting,
namely a Starbucks Coffee House with a cylindrical front wall, on Central Avenue in Albuquerque, New Mexico.  In spite of perturbations to the cylindrical geometry in the form of persons eating bagels, etc., I could hear every word of a conversation at the other side of the room, against a noisy background.

\begin{figure}
\begin{minipage}[b]{.46\linewidth}
\centering\includegraphics[width=40mm]{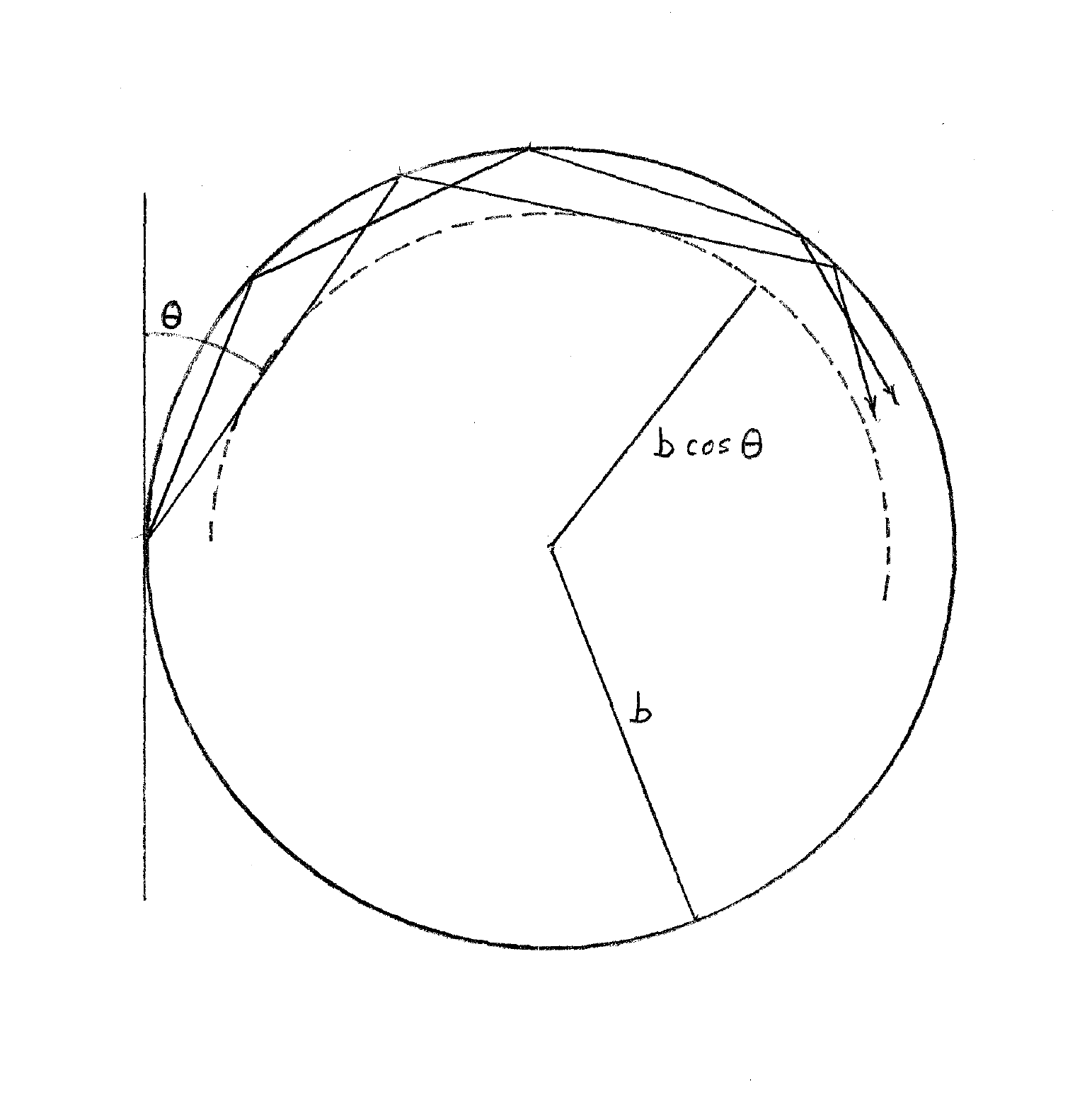}
\caption{Rayleigh's ray picture}
\label{raypic}
\end{minipage}
\begin{minipage}[b]{.46\linewidth}
\centering\includegraphics[width=40mm]{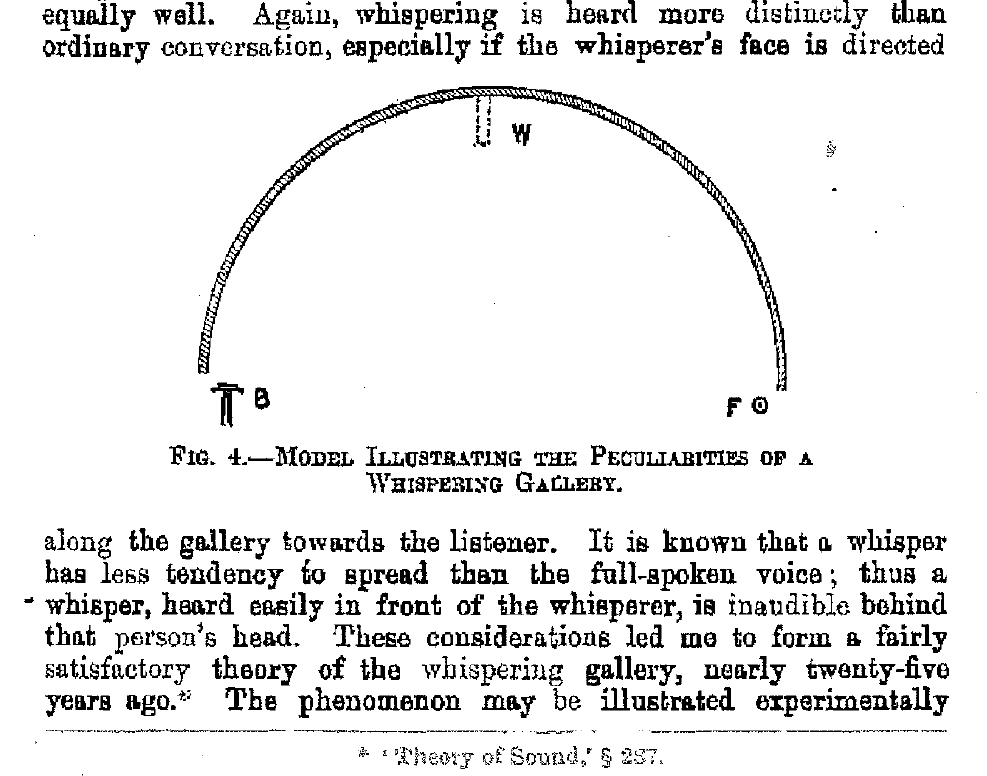}
\caption{Rayleigh's lab experiment}
\label{labexp}
\end{minipage}
\end{figure}

The high frequency components of a speaking voice, prominent in a whisper, are emitted within a narrow cone, say of opening angle $\theta$. Then the minimum distance from an emitted ray to the center of the gallery of radius $b$ is
$b\cos\theta$, as shown in Fig.\ref{raypic}. Thus Rayleigh expected the acoustical disturbance to be localized near the
wall, and performed a charming experiment to verify this ray picture, as in Fig.\ref{labexp}. He set up a sheet of zinc in cylindrical form (2 feet wide by 12 feet long), making an arc of 180 degrees,  and employed a tangentially directed bird call as source and a sensitive flame as detector. The flame flared when the bird call sounded, but the flare could be stopped by imposing a narrow barrier, only 2 inches wide, near the wall. Barriers along the
straight line between whistle and flame had no such effect. In Rayleigh's nicely turned expression,
``Especially remarkable is the narrowness of the obstacle, held
close to the concave surface, which is competent to intercept most of the effect."

Rayleigh went on to develop a wave theory of the phenomenon, in two interesting papers from which we can still learn today \cite{rayleigh_wave}. Let $\psi  $ be the velocity potential of a sound field in a two-dimensional region specified in polar coordinates $(r,\theta)$ with $r\le b$. It satisfies the wave equation
\be
\triangle\psi-\frac{1}{v^2}\frac{\pt^2\psi}{\pt t^2}=0\ .
\ee
An elementary solution  is in terms of the Bessel function $J_n$,
\be
\psi=J_n(kr)\cos(n\theta-kvt)\ .
\ee
This is a wave traveling in the azimuthal direction with frequency $kv$ and wave number $n/R=2\pi/\lambda$.
The boundary condition is that the radial velocity $v_r=\pt\psi/\pt r$  be zero on the boundary $r=b$:
\be
J_n^\prime(kb)=0\quad \Rightarrow \quad kb=j^\prime_{ns},\quad s=1,2,\cdots\ ,
\ee
where the $j\pr_{ns}$ are the  zeros of  $J\pr_n(x)$ enumerated by the integer $s$. The $j\pr_{ns}$ and the corresponding zeros $j_{ns}$ of $J_n(x)$ are all greater than $n$, and are given by asymptotic series at large $n$; see \cite{as}.

This is a resonance condition, satisfied only at discrete frequencies $\omega_*=k_*v$. The solution of the inhomogeneous wave equation with source blows up at those frequencies. One can also think of the resonance condition as a dispersion relation, which relates frequency to wave number at discrete frequencies.

Rayleigh showed that this set-up describes whispering gallery behavior, because the wave function with factor $J_n(k_*r)$
is concentrated at values of $r$ near the boundary, the concentration being more pronounced at large $n$. The mathematics of this concentration, very relevant to the present work, was explored more thoroughly in his second paper.

\section{The Electromagnetic Problem in the Vacuum Chamber of an Accelerator}

\begin{figure}
\centering
\includegraphics[width=40mm]{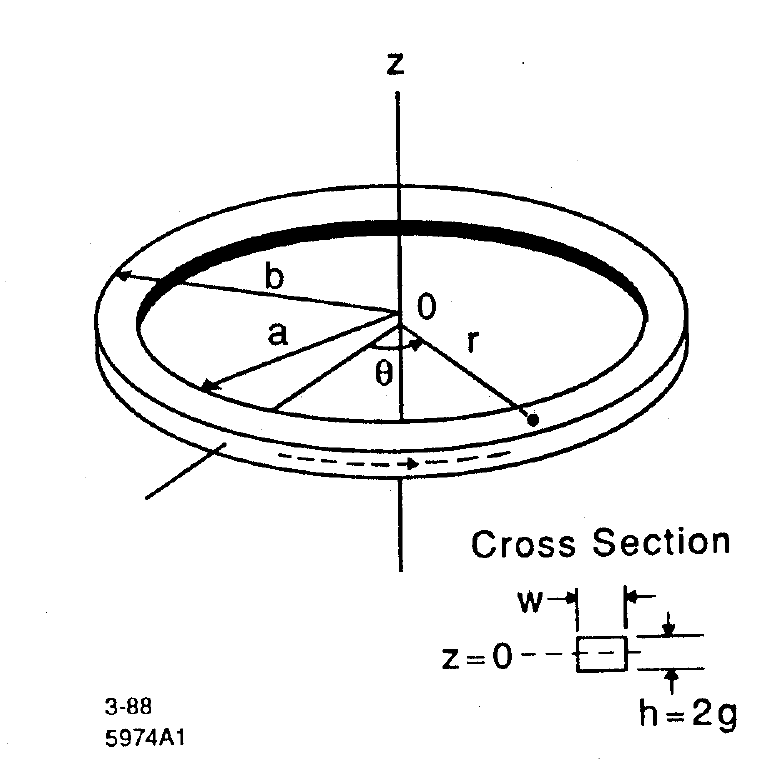}
\caption{Toroidal chamber with rectangular cross section}
\label{toroid}
\end{figure}

Consider a perfectly conducting vacuum chamber of circular toroidal form with rectangular cross section, as illustrated in Fig.\ref{toroid}. After appropriate integral transforms, one can solve Maxwell's equations in the interior of the torus, with correct boundary conditions on the walls, using separation of variables in cylindrical coordinates \cite{wm}.
We take the Laplace transform with respect to time, the variable conjugate to time being $-i\omega$. The complex frequency $\omega$ initially has positive imaginary part $v$, so that convergence of the transform is guaranteed if the field is bounded.  Since the field excited by a circulating beam does not decay at large time, the Fourier transform does not exist. Nevertheless, many papers incorrectly use the Fourier transform, Ref.\cite{wm} among them. We employ Fourier series in $\theta$ and $z$, the latter chosen so as to meet the boundary conditions on the upper and lower planar surfaces of the chamber. For $E_z$ the Laplace-Fourier transform  is
\bea
&&\hat E_z(r,n,p,\omega)=\frac{1}{2\pi}\int_0^{2\pi}d\theta~
e^{-in\theta}\frac{1}{g}\int_{-g}^g dz~\cos\alpha_p(z+g)\nonumber\\
&&\hskip 2.5cm\cdot\frac{1}{2\pi}\int_0^\infty dt~ e^{i\omega t}E_z(r,\theta,z,t)\ ,
\eea
where $\alpha_p=\pi p/h$ with $h=2g$ being the height of the chamber.
All components of the fields can be expressed in terms of $\hat E_z(r,n,p,\omega),\ \hat H_z(r,n,p,\omega)$ and their $r-$derivatives \cite{wm}. The wave equations for $E_z$ and $H_z$ imply that $\hat E_z$ and $\hat H_z$ satisfy Bessel equations with source terms, which can be solved in terms of Bessel functions by variation of parameters \cite{wm}.

The charge-current source can be arbitrary, but to get the simplest possible expression we take a ribbon beam with rigid longitudinal form. Its charge density is
\bea
&&\rho(r,\theta,z,t)=q\lambda(\theta-\omega_0t)H(z)\frac{\delta(r-R)}{R}\ ,\\
&& \int\lambda(\theta)d\theta=\int H(z)dz=1\ ,
\eea
with $H(z)$ an arbitrary vertical profile. The Laplace-Fourier transform of this function is
\bea
&&\hat\rho_{np}(r,\omega)=\frac{iq\lambda_nH_p\delta(r-R)}{2\pi R(\omega-n\omega_0)}\ ,\\
&& \lambda_n=\frac{1}{2\pi}\int_0^{2\pi}e^{-in\theta}\lambda(\theta)d\theta\ ,\quad
H_p= \frac{1}{g}\int_{-g}^g \sin\alpha_p(z+g)H(z)dz\ .
\eea
The corresponding current density is $(J_r,J_\theta,J_z)=(0,\beta cr\rho/R,0)$.

A general charge density,
\be
\rho(r,\theta,z,t)=q\phi(\theta-\omega_0t,r,z,t)\ ,
\ee
has Laplace-Fourier transform
\bea
&&\hat\rho_{np}(r,\omega)=\frac{q}{2\pi}\int_0^\infty e^{i(\omega-n\omega_0)t}\hat\phi_{np}(r,t)dt\ ,\nonumber\\
&& \hat\phi_{np}(r,t)=\frac{1}{2\pi}\int_0^{2\pi}e^{-in\theta}d\theta~\frac{1}{g}\int_{-g}^g\sin\alpha_p(z+g)
~\phi(\theta,r,z,t)dz\ .
\eea
Notice that the $t$-independent part of $\hat\phi_{np}$ again gives a pole at $\omega=n\omega_0$.
 In allowing a  general form of charge-current one must account for the continuity equation, which can be done by constructing charge and current densities from a distribution function in phase space, evolving by correct dynamics.

If the beam is centered in the chamber, the fields at the position of the beam are hardly affected if the radius $a$ of the inner wall is pushed to zero, thus obtaining a ``pill box" chamber rather than a torus. This is not obvious, but comes out of a detailed analysis.
Solving for the longitudinal electric field $E_\theta(r,\theta,z,t)$ in the pill box by the method sketched above we find
its Laplace-Fourier transform as follows (SI units):

\bea
&&\hat E_\theta(r,n,p,\omega)=-\frac{q\beta cZ_0\lambda_nH_p}{4(\omega-n\omega_0)}\nonumber\\
&&\cdot\bigg[\frac{\omega}{c}\bigg(\frac{J_n\pr(\gamma_pr)}{J_n\pr(\gamma_pb)}s_n(\gamma_pb,\gamma_pR)+
\Theta(r-R)s_n(\gamma_pr,\gamma_pR)\bigg)\nonumber\\
&&+\frac{n}{\beta R}\bigg(\frac{\alpha_p}{\gamma_p}\bigg)^2\bigg(\frac{J_n(\gamma_pr)}{J_n(\gamma_pb)}
p_n(\gamma_pb,\gamma_pR)+\Theta(r-R)p_n(\gamma_pr,\gamma_pR)\bigg)\bigg]\label{etheta}\\
\nonumber\\
&&\gamma_p^2=(\omega/c)^2-\alpha_p^2\ ,\quad \alpha_p=\pi p/h\ ,\nonumber\\
&&p_n(x,y)=J_n(x)Y_n(y)-Y_n(x)J_n(y)\ ,\nonumber\\ &&s_n(x,y)=J\pr_n(x) Y\pr_n(y)-Y\pr_n(x) J\pr_n(y)
\eea
Here $Z_0=120\pi\ \Omega$ is the impedance of free space and $\Theta(x)$ is the unit step function,
equal to 1 for $x\ge 0$ and zero otherwise. This formula  displays the same concentration near the outer wall as in Rayleigh's case and similar resonances.
Resonances result from boundary conditions, as in the Rayleigh theory, and give poles in the $\omega$-plane:
\be
J\pr_n(\gamma_pb)=0\quad (TE)\ , \quad J_n(\gamma_pb)=0\quad (TM)\ ,\label{res}
\ee
In our nomenclature the TE and TM modes have electric and magnetic fields transverse to the symmetry axis of the problem ($z$-axis). Thus the TE mode has electric field polarized in the plane of motion of the beam, while the TM mode has electric field perpendicular to that plane.

\begin{figure}
\begin{minipage}[b]{.46\linewidth}
\centering\includegraphics[width=40mm]{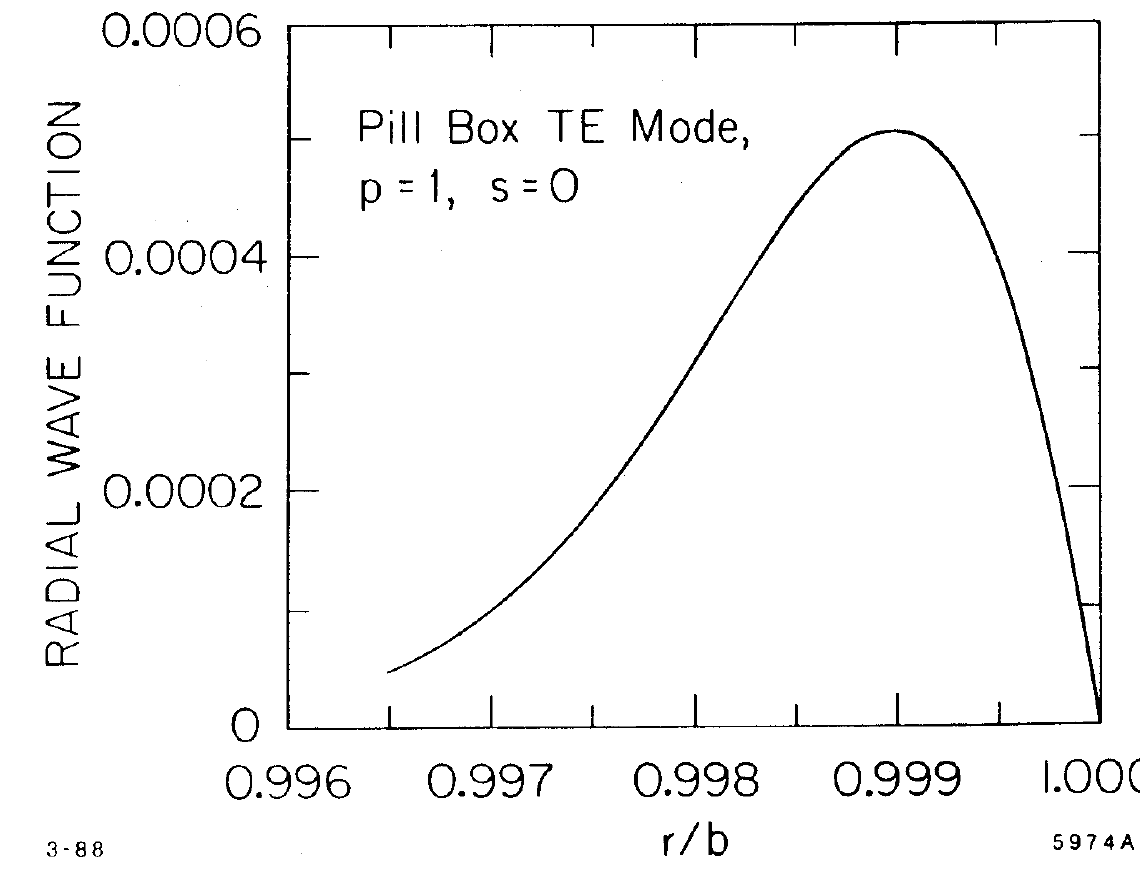}
\end{minipage}
\begin{minipage}[b]{.46\linewidth}
\centering\includegraphics[width=40mm]{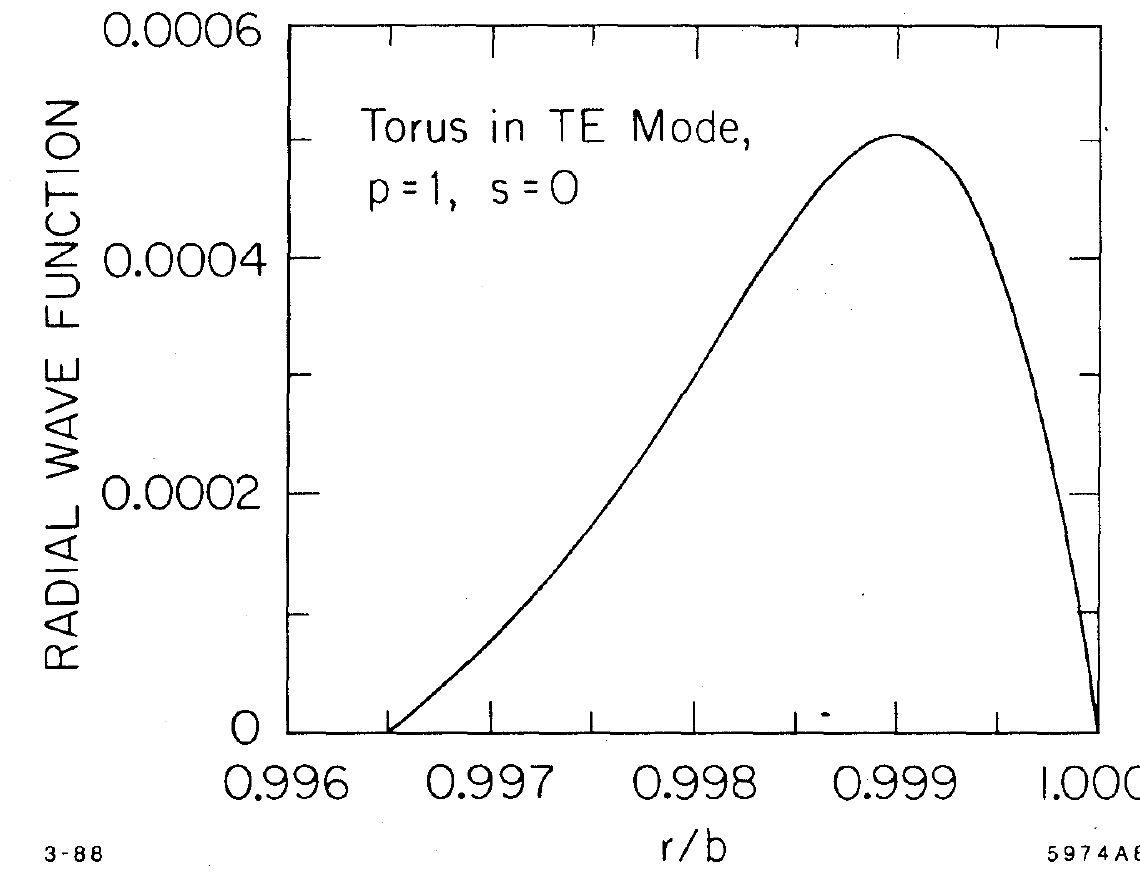}
\end{minipage}
\caption{Wave functions vs. $r/b$ for pill box and torus. Here $s=0$ corresponds to $s=1$ in the main text.}
\label{wavefcn}
\end{figure}

As remarked above the toroidal and pill box models give nearly the same field at the position of the beam, and indeed at any point not too close to the inner torus wall.
The corresponding wave functions for a typical choice of parameters are shown in Fig.\ref{wavefcn}.

The longitudinal field (averaged over transverse distributions) can be expressed in terms of the impedance
$Z(n,\omega)$:
\be
-2\pi R{\mathcal E}_\theta(n,\omega)=Z(n,\omega) I(n,\omega)\ ,
\quad  I(n,\omega)=\frac{iq\omega_0\lambda_n}{2\pi(\omega-n\omega_0)}
\ee
The wake voltage is given by the inverse Laplace-Fourier transform,
\be
V(\theta,t)=\sum_ne^{in\theta}\int_{\Imp\omega=v}e^{-i\omega t}Z(n,\omega)I(n,\omega)d\omega\ ,\quad v>0\ ,\label{wakev}
\ee
and the power is ${\mathcal P}=-dW/dt$, where $W$ is the work done per unit time by the wake field,
\be
{\mathcal P}(t)=q\omega_0\sum_ne^{in\omega_0t}\lambda_n^*\int_{\Imp\omega=v} e^{-i\omega t}Z(n,\omega)I(n,\omega)d\omega\ \label{power_exact}
\ee
The impedance deduced from (\ref{etheta}) is
\bea
&&Z(n,\omega)=i\pi^2Z_0gR\sum_{p=0}^\infty H_p^2 \biggl[ \frac{\omega R}{c}\frac{J_{|n|}\pr(\gamma_pR)}
{J_{|n|}\pr(\gamma_pb)}s_{|n|}(\gamma_pb,\gamma_pR)\nonumber\\
&&\hskip 3.1cm +\frac{n}{\beta}\biggl(\frac{\alpha_p}{\gamma_p}\biggr)^2\frac{J_{|n|}(\gamma_pR)}
{J_{|n|}(\gamma_pb)}p_{|n|}(\gamma_pb,\gamma_pR)\biggr]\ .
\eea
This is correct at positive and negative $n$ alike.

Now the wake voltage (\ref{wakev}) may be evaluated by pushing the $\omega$-contour to infinity in  the lower half-plane,
and in so doing one encounters poles at  $\omega=n\omega_0$ from the factor in the current, at resonant frequencies
$\omega=\pm\omega_r$ defined by Eqs.(\ref{res}), and at wave guide poles $\omega=\pm\alpha_pc$ which appear in the
impedance. The contour at infinity in the lower half plane gives no contribution. Thus we have \cite{erber}
\bea
&&V(\theta,t)=\sum_ne^{in\theta}\int_{\Imp\omega=v} e^{-i\omega t}Z(n,\omega)\hat I(n,\omega)d\omega\nonumber\\
&&=q\omega_0\sum_n\lambda_n Z(n,n\omega_0)e^{in(\theta-\omega_0t)}\nonumber\\
&&+q\omega_0\sum_n e^{in\theta}\lambda_n\sum_j\bigg[\frac{e^{-i\omega_jt}\textsf{R}(n,\omega_j)}{\omega_j-n\omega_0}
-\frac{e^{i\omega_jt}\textsf{R}(n,-\omega_j)}{\omega_j+n\omega_0}\bigg]\nonumber\\
&&=V_1(\theta-\omega_0t)+V_2(\theta,t)\ . \label{Vcorrect}
\eea
Here the (positive) resonance and wave guide pole positions are denoted generally by $\omega_j$, and $\textsf{R}(n,\omega_j)$ is the residue of the pole in $Z(n,\omega)$ at $\omega=\omega_j$.

The first term $V_1$ in(\ref{Vcorrect}) depends only on the angular distance $\theta-\omega_0t$ from the reference particle and is the usual expression of the induced voltage that one finds in the literature, with $Z(n,n\omega_0)$ being a more precise designation of what is usually called $Z(n)$. The second term $V_2(\theta,t)$ is conceptually important, but as far as I know does not appear in the literature before Ref.\cite{erber}. It exactly cancels
the first term $V_1$ if $n\omega_0$ should approach one of the resonance frequencies $\omega_r$, where $Z(n,n\omega_0)$
is infinite. Without the cancelation we have a potentially infinite induced voltage, which hardly seems physical.

When wall resistance is included in the boundary condition, as was done in Ref.\cite{wm}, the resonance poles are displaced to nearby points in the lower half-plane: $\pm\omega_r\rightarrow \pm\omega_r-i\epsilon$. Then the corresponding terms in $V_2$ vanish at large $t$ with a damping factors $\exp (-\epsilon t)$. There is also a third term $V_3(\theta,t)$
arising form a branch point of the impedance at $\omega=0$ due to the square root singularity in the skin depth.
This too vanishes at large $t$, being an integral along the negative imaginary axis with factor $\exp(-i\omega t)$ in the integrand.
It appears that the contribution of the wave guide poles to $V_2$ averages to zero with increasing $t$, but a careful analysis of this point remains to be done.

We invoke wall resistance and carry out a similar treatment of the power (\ref{power_exact}). Furthermore, we suppose that when
the bunch form evolves in time, the constant $\lambda_n$ can simply be replaced by $\lambda_n(t)$. As was shown in Refs.\ \cite{deform1} and \cite{deform2} for the case of the parallel plate model of the vacuum chamber, this is only an approximation, and not always an accurate one.  A careful study of the approximation for both parallel plates and the toroidal model would be desirable. For now we just declare the time-dependent induced voltage and power to be
\bea
&&V(\theta,t)=q\omega_0\sum_{n=-\infty}^\infty e^{in(\theta-\omega_0t)}Z(n,n\omega_0)\lambda_n(t)\ ,\label{Vt}\\
&&P(t)=2(q\omega_0)^2\sum_{n=1}^\infty\Rep Z(n,n\omega_0)|\lambda_n(t)|^2\ .\label{Pt}
\eea

\section{Comparison with Spectrum Measured at NSLS-VUV Light Source}
\begin{figure}[htbp]
\begin{center}
\includegraphics[width=3.5in]{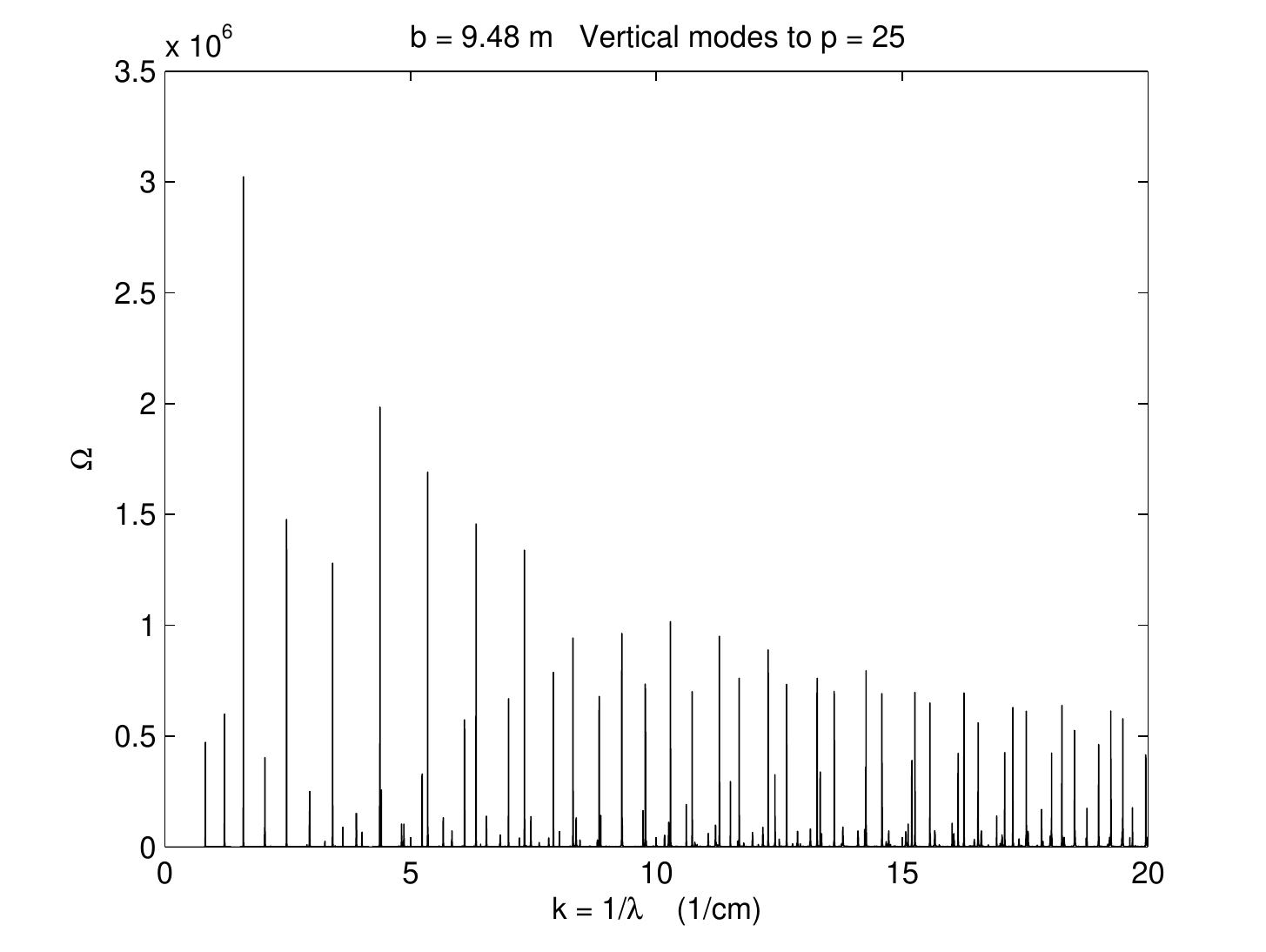}
\vspace{-.3cm}
\caption{$\Rep Z(n,n\omega_0)$ for parameters of VUV light source, vs. wave number $1/\lambda$ in
units of cm$^{-1}$.}
\label{ZVUV}
\end{center}
\end{figure}
\begin{figure}[htbp]
\begin{center}
\includegraphics[width=3.5in]{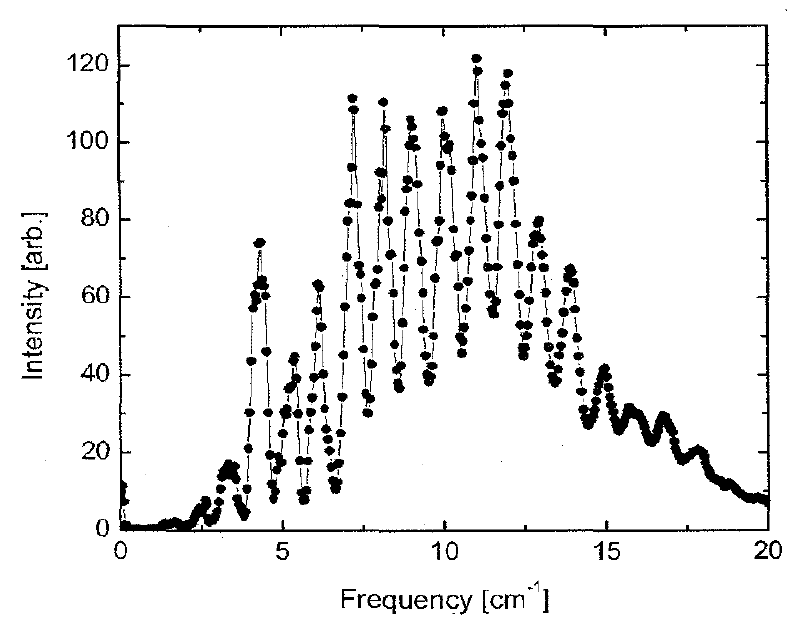}
\vspace{-.75cm}
\caption{Far IR spectrum measured at NSLS}
\label{VUVhi}
\end{center}
\end{figure}
\begin{figure}[htbp]
\begin{center}
\includegraphics[width=3.5in]{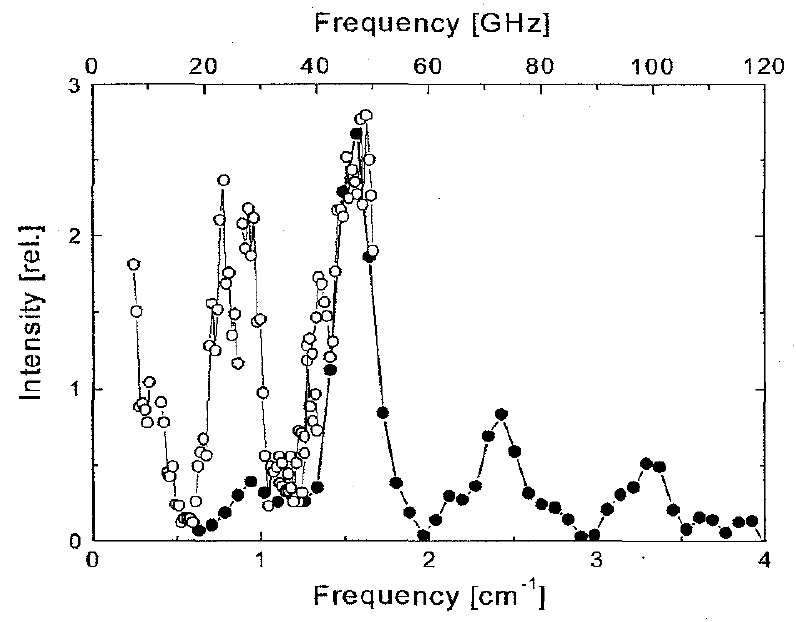}
\vspace{-.5cm}
\caption{Low frequency  spectrum measured at NSLS. The black dots are IR data from
an interferometer, the open dots from RF measurements.}
\label{VUVlo}
\end{center}
\end{figure}

\begin{table}

\begin{center}
\caption{Theoretical frequencies compared to data of Figs.\ref{VUVhi},\ref{VUVlo} }
\vskip .2cm
\setlength{\tabcolsep}{.25in}
\begin{tabular}{cc|cc}

\textbf{Exp.} & \textbf{Thy.} & \textbf{Exp.} & \textbf{Thy.}\\ \hline
0.80& 0.827& 6.10&6.31\\
0.93&--- &7.25&7.32\\
1.32&1.21&9.00&8.32\\
1.57&1.60&10.0&9.29\\
2.10*&2.04&11.1&10.28\\
2.40&2.48&12.0&11.29\\
2.76*&2.94&12.8&12.33\\
3.10*&3.26&13.8&13.31\\
3.66*&3.62&15.0&14.3\\
3.88*&3.90&15.7-15.9&15.3\\
4.20&4.38&16.7&16.3\\
5.25&5.34&18.0 &17.3\\
& &18.8*&18.3
\end{tabular}
\end{center}
\end{table}

The vacuum chamber in the NSLS-VUV synchrotron light source at Brookhaven has relatively little deviation from the toroidal form
with rectangular cross section, in the bending magnets where CSR is observed. Ignoring the effect of straight sections and chamber corrugations elsewhere in the ring we make a comparison of the observed CSR power spectrum to the real part of the impedance computed from the toroidal model with resistive wall, shown in Fig.\ref{ZVUV}. The parameters are $R=1.91$m, $w=8$cm, $h=4$cm,
where $R$ is the bending radius and  $w$ and $h$ are  the horizontal width and vertical height of the chamber. Vertical modes up to $p=25$ are included. The wall conductivity is that of the stainless steel chamber, but that is not a critical parameter. The outer wall of the torus is at $r=b=1.948$m, and the beam is at $r=R$, putting it $2$mm off center; this improves somewhat the fit to the data in comparison to a centered beam. Measurements \cite{Carr} were done with a Michelson interferometer (Fig.\ref{VUVhi}) and, at the lowest frequencies near the cut-off of the radiation impedance, with RF methods
(horn antenna, wave guides, and frequency analyzer) (Fig.\ref{VUVlo}). For details of our fit see \cite{PAC11}. Here we show only the comparison of experimental and theoretical lines in Table 1. The entries marked with an asterisk correspond to somewhat doubtful small shoulders in the data, rather than clear peaks.

In a follow up to this fit, D. Zhou \cite{zhou} has done a calculation including straight sections, but without
imposing periodicity of fields around the ring. Periodicity is hard to impose in the paraxial approximation that Zhou employed. Peaks in the impedance in this calculation have non-zero widths, which Zhou tries to identify with experimental widths. I am skeptical of this identification for two reasons. First, I suspect that periodicity will imply sharp resonances, even in the presence of bends; second, there are eminent sources of experimental widths, such as dispersion
in the IR beam lines.  Nevertheless, Zhou raises an interesting issue which certainly should be explored.

\section{High Resolution CSR Spectra at the \break Canadian Light Source (CLS)}

\begin{figure}[htb]
   \centering
  \includegraphics*[width=70mm,height=21.5mm]{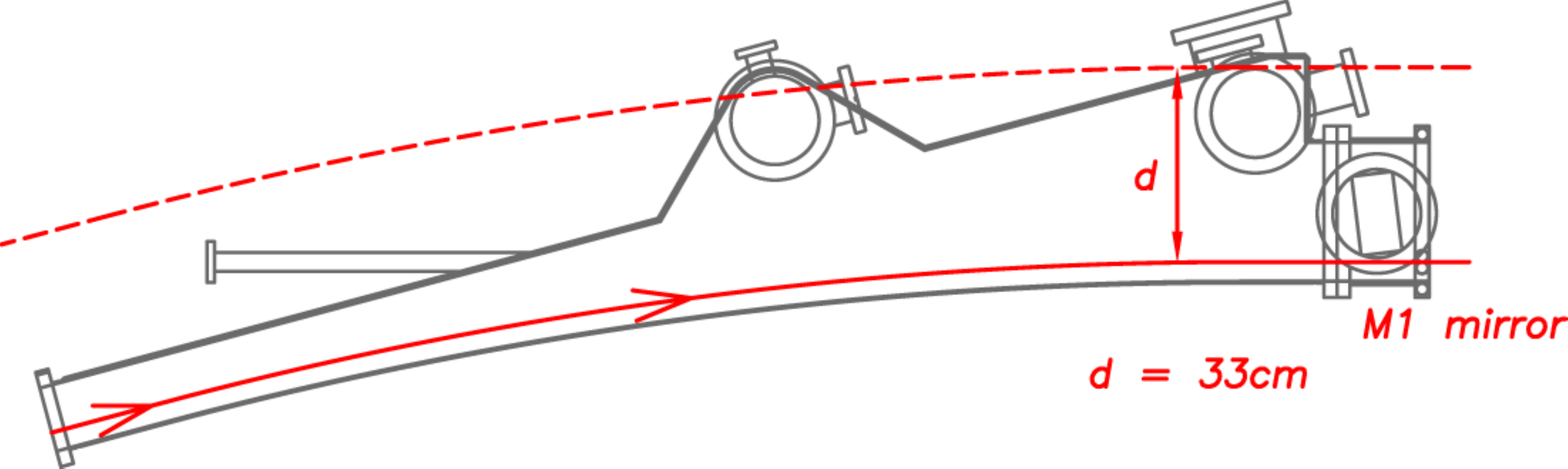}
   \caption{Vacuum chamber of CLS at dipole where IR is extracted}
   \label{CLSchmbr}
\end{figure}

In view of the impressive agreement between theory and the data from VUV, taken a long time ago, one is encouraged
to look at more recent data and other machines. On the one hand, there are now superb instruments such as the Bruker
IFS 125 HR interferometer at CLS and Soleil, with a resolution down to 0.0009 ${\rm cm}^{-1}$. On the other hand,
comparison with theory is more difficult at machines more modern than the VUV, since their vacuum chambers at the location where
IR is extracted do not have the simple form of the toroidal model with constant rectangular cross section. Rather,
there is usually a fluted chamber  with an outer wall receding from the beam by a large distance, perhaps tens of
centimeters. Fig.\ref{CLSchmbr} shows the plan of one of two similar IR chambers at the CLS. Within the chamber are two significant metallic structures that can reflect radiation, a copper photon absorber, and a mirror near the beam to send radiation into the IR beam line leading to the interferometer. We can hardly expect the simple toroidal model to describe fields in this structure, and indeed experiment indicates that it does not.

\begin{figure}[htb]
   \centering
    \includegraphics*[width=70mm,height=35mm]{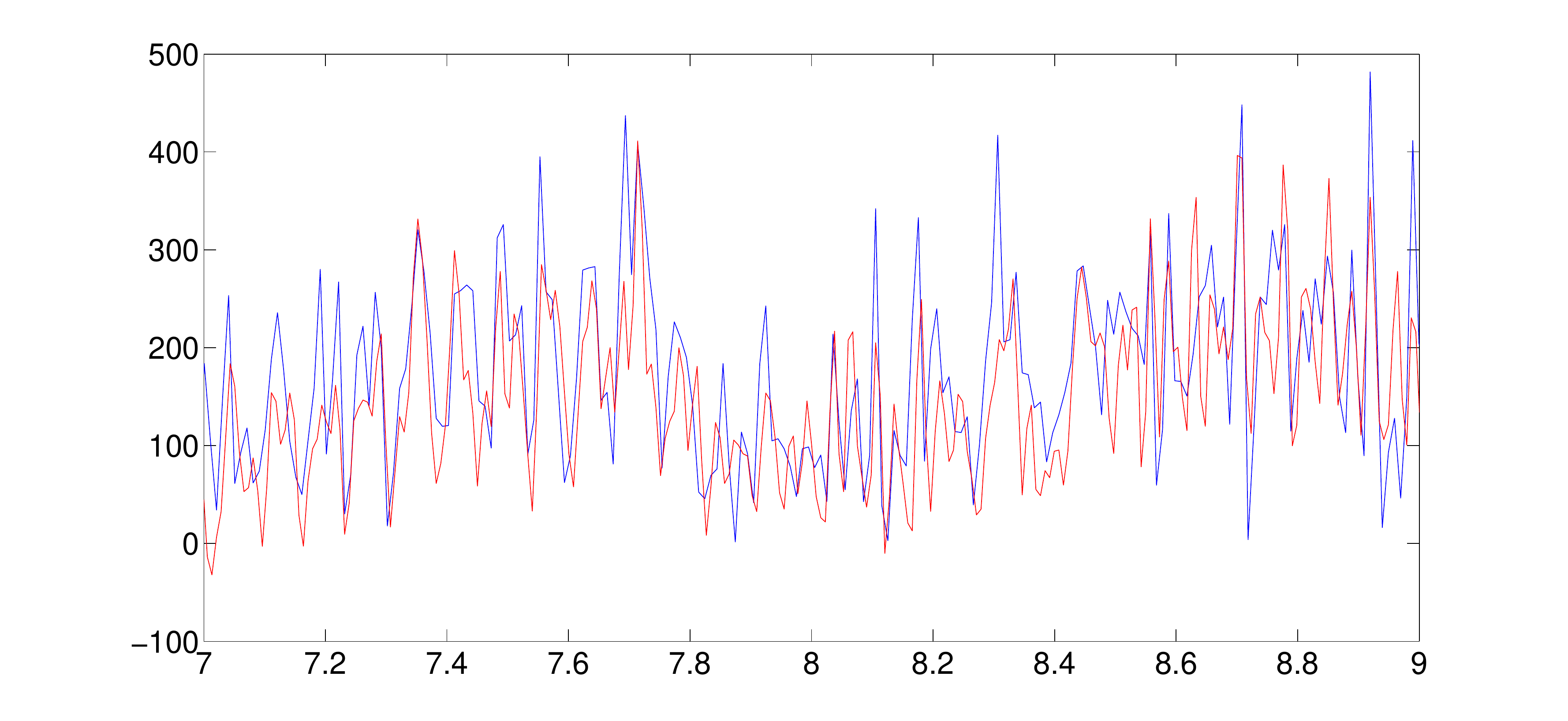}
   \caption{CLS spectra from two very different runs, power (a.u.) vs. $k$ in cm$^{-1}$.}
   \label{CLSspec}
\end{figure}

Nevertheless, observations at the CLS leave little doubt that there are sharp peaks in the power spectrum, which are determined by the vacuum chamber and the bending field alone.  Spectra taken with the Bruker show a remarkable stability with respect to changes in the machine setup and the structure of the IR beam line. Figure~\ref{CLSspec} compares a power spectrum (red) taken in the bursting mode of CSR with one bunch at 2.9 GeV, on 18/5/2010, with another (blue) taken in the continuous mode with 210 bunches at 1.5 GeV, on 30/1/2012. The latter was multiplied by a factor of $8$ to aid comparison. Because of the large change in the beam one expects the positions of peaks to agree better than the relative heights, as is found, but even the relative heights show a lot of similarity between the two runs. We take this stability as a strong indication that the spectrum is determined primarily by the vacuum chamber and  bends.

The peaks in the fine structure of Fig.\ref{CLSspec} have a spacing   $\Delta k =\Delta(1/\lambda)\approx 0.073 {\rm cm}^{-1}$.
If we try to get that spacing from the toroidal model,
the distance from the beam to the outer wall must be $d=b-R=33{\rm cm}$, which happens to be near the actual distance at the maximum excursion of the wall. Here the spacing referred to is that of the strongly dominant TE modes.

\section{Implications for the Wake Field}
Suppose there is only one bunch in the ring. The wake field within the confines of the bunch  turns out to be almost the same for the toroidal and parallel plate impedance models, at least for a Gaussian bunch that is short compared to the bending radius. This in spite of the utterly different appearance of the two impedances.  This behavior was noticed long ago \cite{bobkarl} and unfortunately led to my opinion that the whispering gallery modes would have no great influence on bunch dynamics.

\begin{figure}[htb]
   \centering
  \includegraphics*[width=65mm]{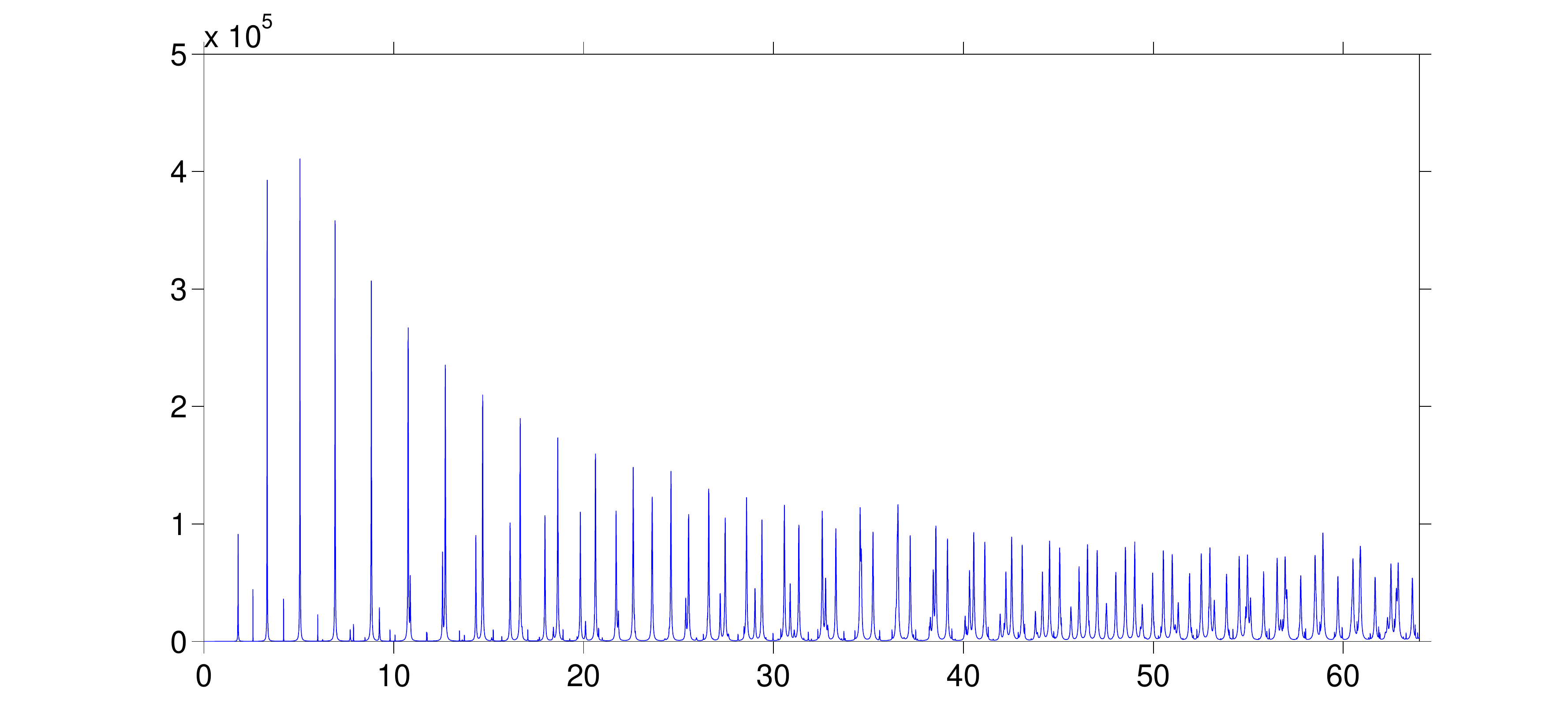}
   \caption{Re $Z(k)$ in ohms for ANKA parameters, vs.
     $k = 1/\lambda\ $ in ${\rm cm}^{-1}$.}
   \label{ZANKA}
\end{figure}

\begin{figure}[htb]
   \centering
     \includegraphics*[width=55mm,height=25mm]{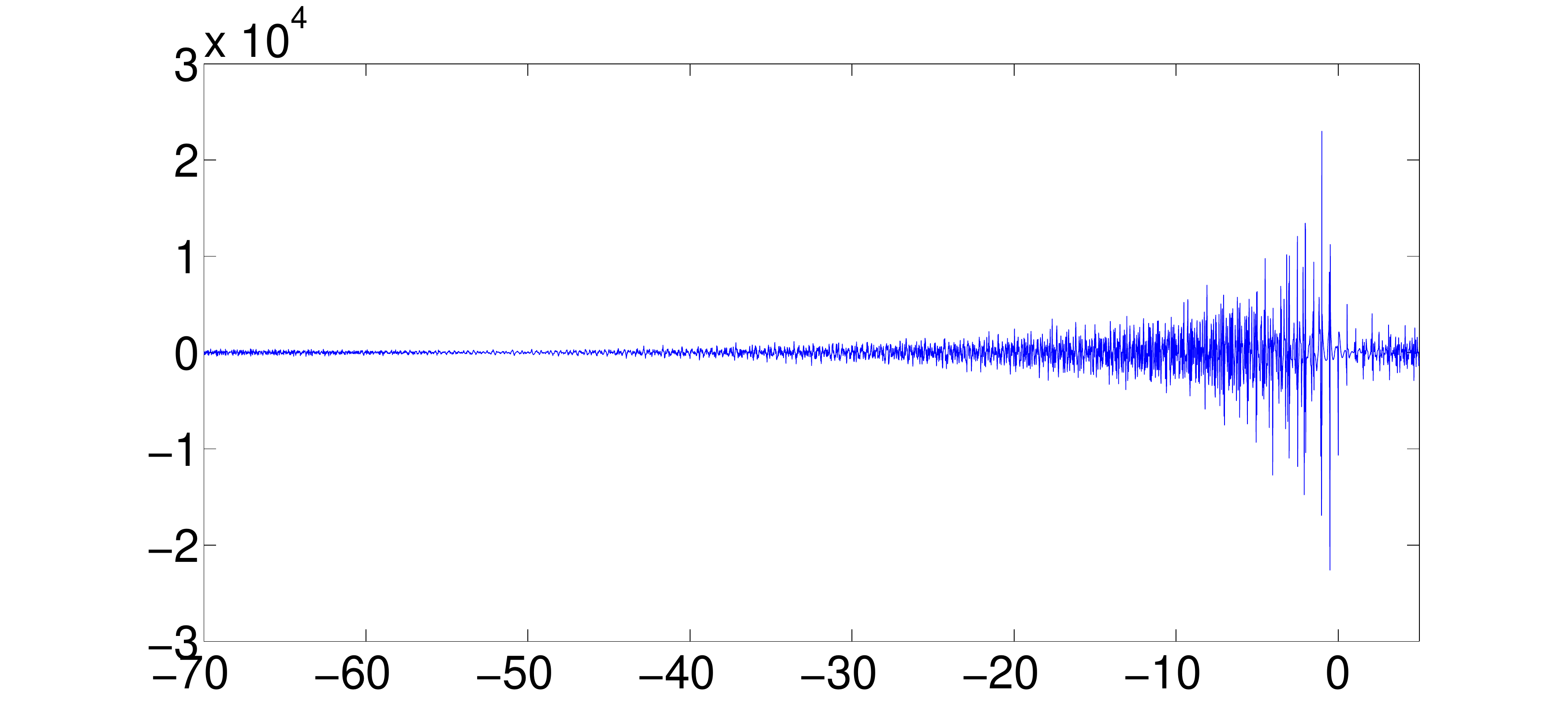}
    \caption{Wake potential $W(z)$ in V/pC from  $Z$ of Fig.\ref{ZANKA}\ , vs. $z$ in cm.}
   \label{wakeANKA}
\end{figure}

When there are two or more bunches in the ring and the bunch currents are large the situation is quite different.
The toroidal model predicts a very long wake field, which can affect the dynamics of a following bunch. In Fig.\ref{ZANKA} we show the real part of the toroidal impedance for parameters of the ANKA light source in Karlsruhe. The corresponding wake potential, computed from (\ref{Vt}) with $\lambda_n=1/2\pi$ and maximum $n$ corresponding to $k= 65 {\rm cm}^{-1}$, is shown in Fig.\ref{wakeANKA}; the head of the bunch is on the right. A feature of the model, if not of the real system, is that the wake wraps all the way around the ring, so that there is a precursor field in front of the bunch.

\section{Experimental Evidence for Interbunch \break Communication}
\begin{figure}[htb]
   \centering
     \includegraphics*[width=55mm,height=25mm]{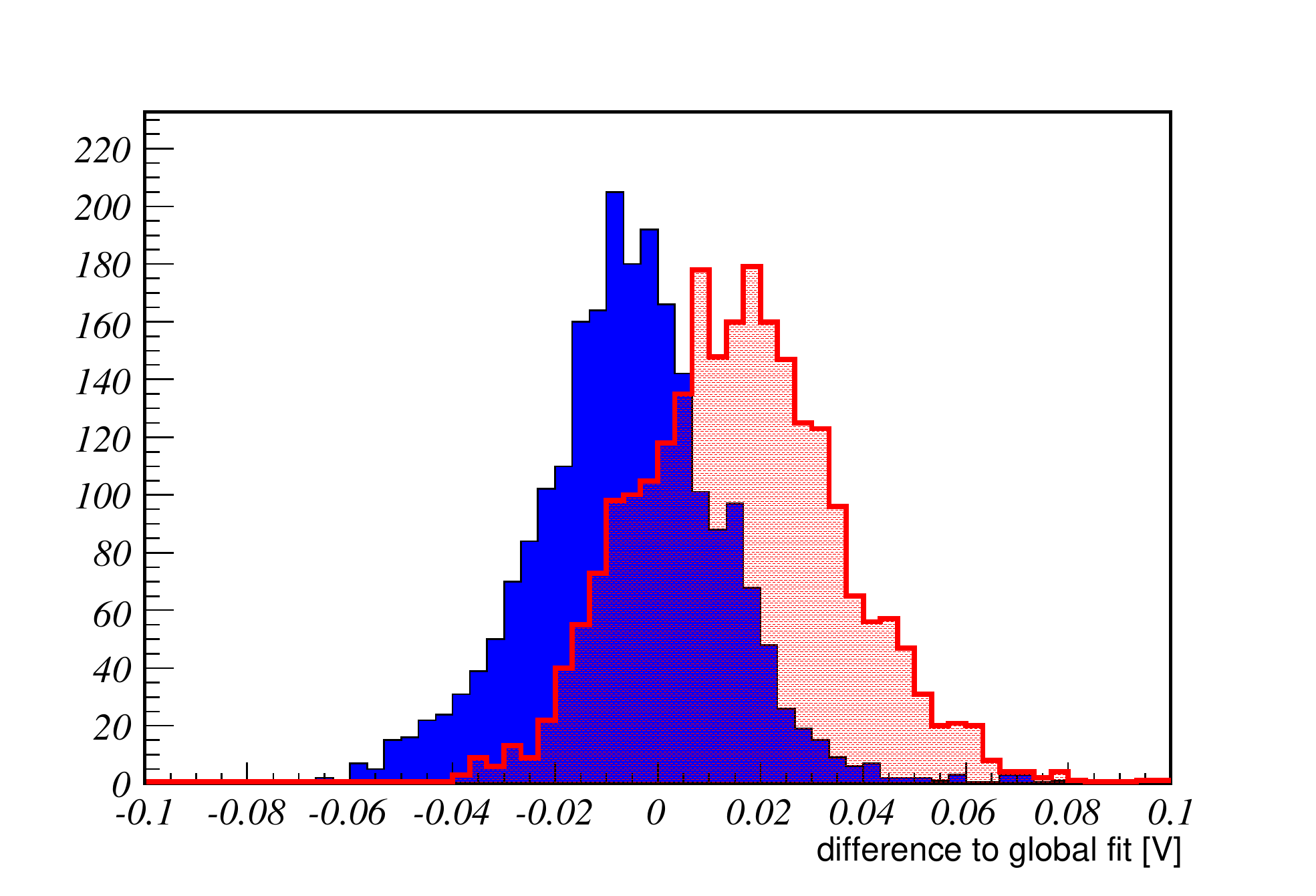}
   \caption{Radiation enhancement by higher charge in preceding bunch.}
   \label{Judin}
\end{figure}

 Evidence of a long range wake field comes from an experiment at ANKA by V. Judin  and collaborators \cite{judin},\ \cite{judin-phd}.
They observed THz radiation with a fast bolometer having sufficient time resolution to distinguish radiation from individual bunches. A large number of buckets were filled with known but varying amounts of charge. Bolometer signals from the various bunches were sorted into two groups: the blue group in which the preceding bunch in the fill has at least 10\% less charge, and the red group in which the preceding bunch has at least 10\% greater charge. The power signals were preponderantly greater for the red group, as is seen in the histogram of Fig.~\ref{Judin}. The histogram gives the distribution of red and blue signals relative to a curve which is a global  fit to all the signals.

\section{Direct Observation of a Long Range Wake Field}
\begin{figure}[htb]
   \centering
      \includegraphics*[width=55mm,height=25mm]{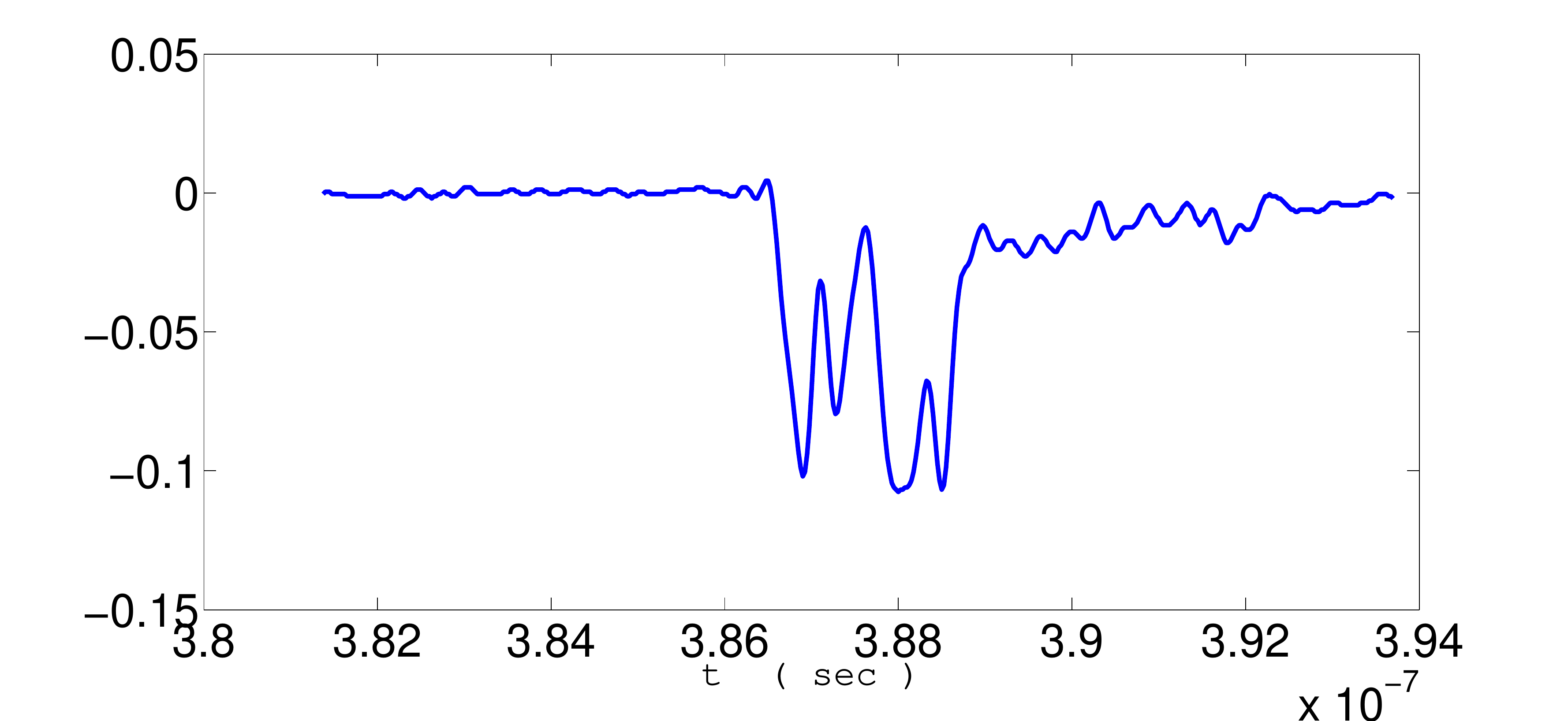}
   \caption{Signal (a.u.) from backward viewing diode at CLS vs. time in seconds.}
   \label{backport}
\end{figure}

Experiments which aspire to observe the wake field directly were carried out at the CLS, following an idea of S. Kramer to put a microwave horn and diode detector at a backward port (in the horizontal pipe seen in Fig.~\ref{CLSchmbr}) in one of the two IR dipole chambers. The diode has a bandwidth of roughly 50-75 GHz, and receives signals traveling opposite to the direction of the beam, which might come from backward reflections off structures present in the chamber.
Fig.~\ref{backport} shows a typical oscilloscope trace of the diode signal. Labeling the prominent {\it downward} peaks from left to right as 1 to 4, we have a plausible explanation as follows: 1 and 2 are reflections from the first downstream obstacle in the flared chamber, a copper photon absorber (in second circle from right in Fig.~\ref{CLSchmbr}), whereas 3 and 4 are from the next obstacle, a structure supporting the M1 pick-off mirror. The 1-3 separation corresponds closely to the separation of these obstacles. Peak 1 is seen as the prompt wake field from the bunch, while peak 2 is a delayed pulse in the wake field, about $13.5\ {\rm cm}$ behind the bunch; similarly for 3 and 4, coming from the later reflection. The point to emphasize is that the distance from bunch to delayed wake pulse is very close to the reciprocal of the average spacing of peaks in the power spectrum shown in Fig.~\ref{CLSspec}, namely $\Delta k\approx 0.073 {\rm cm}^{-1}$. Correspondingly, the distance between the center burst and the first side peak in the interferogram is $13.5$cm.

The interpretation of the peaks in terms of reflections is given added weight by a second experiment in which the diode was moved to a ``normal" dipole chamber. That resembles the special IR chamber, but lacks the pick-off mirror and has a slotted wall (centered slot of width 1 cm) between the beam and the large flared box. The analog of peaks 1-2 is seen, but that of 3-4 goes away, in accord with the absence of the mirror support structure.

\section{Simulation of Interbunch Communication}

In collaboration with Marit Klein, I have
 demonstrated the effect of the long range wake field from whispering gallery modes by solving a nonlinear
  Vlasov-Fokker-Planck (VFP) system for two bunches in two adjacent buckets. We take ANKA parameters for which the interbunch spacing is $60$cm. The impedance is that of Fig.~\ref{ZANKA}.  We have two coupled VFP equations, each referring to the longitudinal phase space distribution for a bunch in its beam frame, but with a term in the wake field defined by the distribution of the other bunch. The
equations are solved by the method of Ref.\cite{bobjim}
, but with bi-cubic rather than bi-quadratic interpolation to update the distribution. The initial distributions are Ha\" issinski equilibria, which are highly unstable at the currents considered.

\begin{figure}
 \centering
     \includegraphics*[width=65mm,height=30mm]{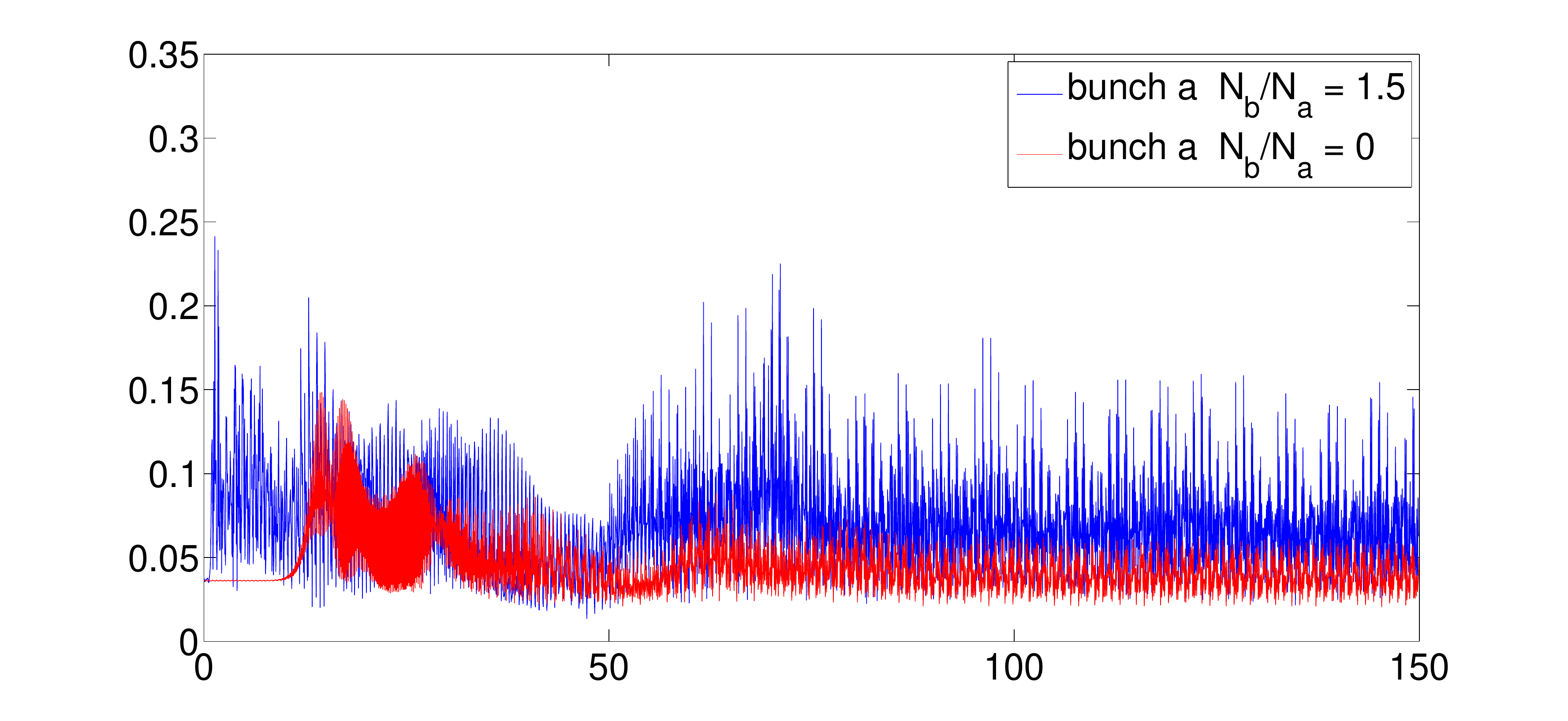}
   \caption{Power from two bunches with equal charges.}
   \label{fig:power-a-and-b}
 \end{figure}

 We plot the total coherent power (a.u.) radiated by each of the two bunches vs. time in synchrotron periods.  There are $N_a=1.14\cdot 10^9$ particles ($0.49$mA )in bunch $(a)$, and an unperturbed bunch length for both bunches of $\sigma_z=1.92$mm, typical for a low-$\alpha$ setup of ANKA used in CSR runs. The longitudinal damping time is given its realistic value. For Fig.~\ref{fig:power-a-and-b} we have $N_a=N_b$, but  the power from the trailing bunch (blue) is consistently greater than that from the leading bunch (red).

\begin{figure}
 \centering
   \includegraphics*[width=65mm,height=30mm]{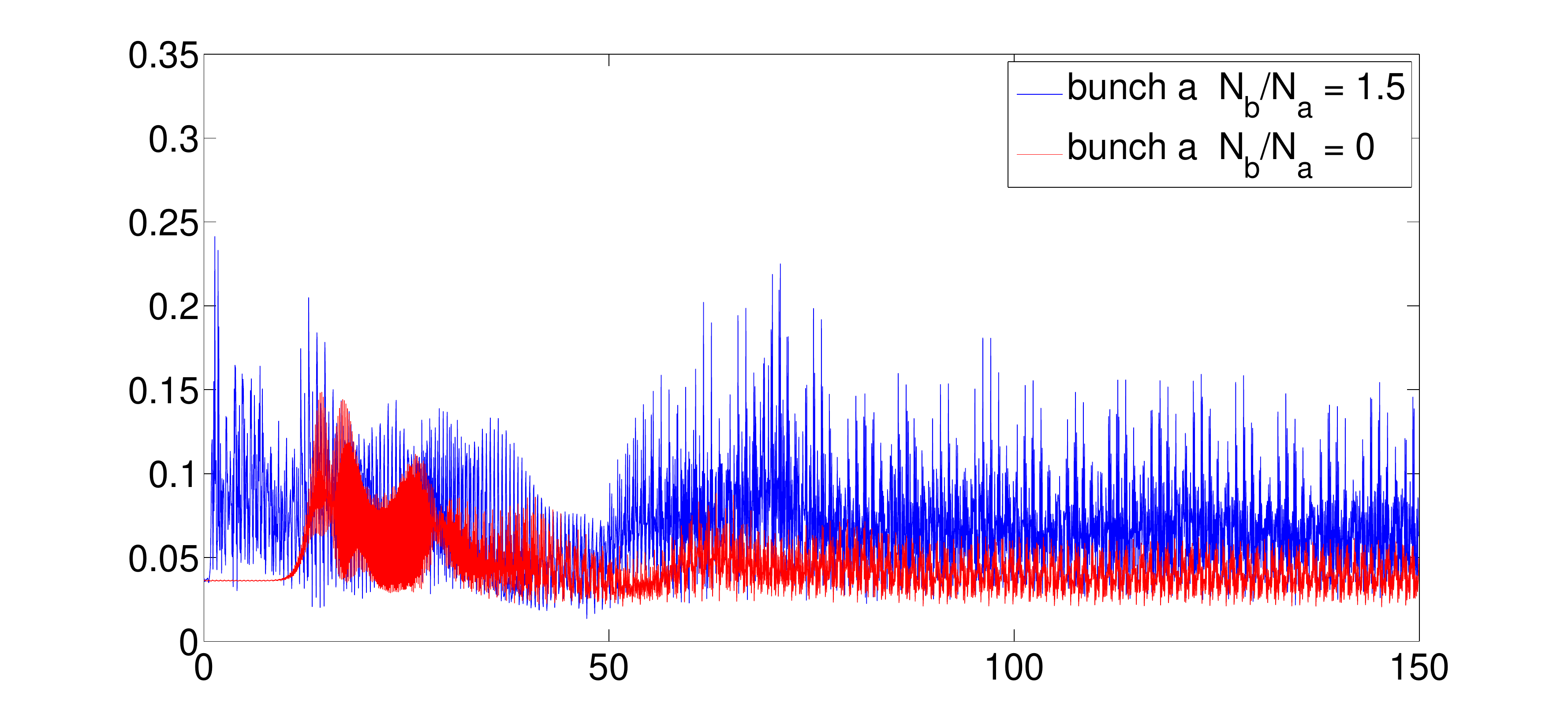}
   \caption{Power from trailing bunch (a), with and without leading bunch, which has 50\% more charge.}
   \label{fig:power-a-vs-rba}
 \end{figure}

 Fig.~\ref{fig:power-a-vs-rba} shows the power from trailing bunch (a), without a leading bunch (red) and with a leading bunch having 50\% more charge (blue). The strong enhancement due to the leading bunch is perhaps surprising in view of the seemingly small   wake potential at $60$cm shown in Fig.~\ref{wakeANKA}, but is believed to be an authentic consequence of the model, evidently a feature of the unstable bunch dynamics at the large (but realistic) currents considered. Of course, the corresponding calculation with the parallel plate impedance shows no inter-bunch communication.

\section{Outlook}
I have reviewed the prediction of sharp peaks in the power spectrum of CSR on the basis of an idealized model of the
vacuum chamber,  and have compared the prediction to  experiment. For more precise comparisons, one will need a theory
accounting for the specific geometric shape of the vacuum chamber in the bend region where radiation is produced and sent to the spectrometer. It will be important to understand the question of locality of the phenomenon: does it depend in part on fields produced in upstream bends, or not?  I hope that these issues will be clarified by a new frequency domain theory which
takes a global view of a closed vacuum chamber with arbitrary bends and straights and arbitrary outer wall excursions.

On the experimental side there are ongoing experiments at the CLS, using backward and forward viewing diodes, sensitive to polarization, together with the Michelson interferometer. With more precision and a greater range of tests, a consistent picture seems to be emerging. It is hoped that a better understanding might lead to a way of smoothing out peaks in the spectrum, which are an annoyance to users of the IR facility.

I have not touched on the implications for bunch dynamics, beyond showing that radiation from a trailing bunch is enhanced compared to that of a leading bunch. Dynamical questions clearly need more theoretical and experimental study. Simulations with a large number of bunches are feasible with parallel computation, as has been shown at SOLEIL. Among other efforts one tries to reproduce ``waterfall plots" obtained experimentally, that is to say the Fourier transform of the radiated power with respect to time, versus the current, the latter naturally declining during a fill.

  For simulation of bunch dynamics it may be less important to understand the fields in the fluted IR chambers, since the average wake field over a turn is all that we need, one turn being much less than a synchrotron period. The chamber is simpler in most of the bends, and much easier to model.

\section{Acknowledgment}
I am forever indebted to Jack Bergstrom of the Canadian Light Source, who has inspired and encouraged this work over many years. Marit Klein contributed to recent calculations
and interesting discussions. On the experimental side, Steven Kramer, Brant Billinghurst, Vitali Judin, and many others did invaluable work. Partial support of work at SLAC was from U.S. Dept. of Energy contract  DE-AC03-76SF00515; support at CLS was from CFI, NSERC, NRC, The Canadian Institutes of Health Research, and the Government of Saskatchewan.

\end{document}